\documentclass[aps,twocolumn,prb,showpacs,superscriptaddress]{revtex4-2}
\usepackage{makeidx}
\usepackage{amsfonts}
\usepackage{graphics}
\usepackage{graphicx}
\usepackage{amssymb}
\usepackage{amsthm}
\usepackage{amsmath}
\usepackage{hyperref}

\begin{document}

\title{Effect of interatomic repulsion on Majorana zero modes 
in a coupled quantum-dot-superconducting-nanowire hybrid system}

\author{R. Kenyi Takagui P\'{e}rez}
\affiliation{Centro At\'{o}mico Bariloche and Instituto Balseiro, 8400 Bariloche, Argentina}

\author{A. A. Aligia}
\affiliation{Instituto de Nanociencia y Nanotecnolog\'{\i}a 
CNEA-CONICET, GAIDI,
Centro At\'{o}mico Bariloche and Instituto Balseiro, 8400 Bariloche, Argentina}

\begin{abstract}
We study the low-energy eigenstates of a topological 
superconductor wire modeled by a Kitaev chain, which is connected at one of its ends to a quantum dot through nearest-neighbor (NN) hopping and NN Coulomb repulsion.
Using an unrestricted Hartree-Fock approximation
to decouple the Coulomb term, we obtain that the quality
of the Majorana end states is seriously affected
by this term only when the dependence of the
low-lying energies with the energy of the quantum dot shows 
a ``diamond'' shape, characteristic of short wires.
We discuss limitations of the simplest effective models to 
describe the physics.
We expect the same behavior in more realistic models for
topological superconducting wires.

\end{abstract}

\maketitle


\section{Introduction}

\label{intro}

In recent years, topological superconducting wires 
has been a field of intense research in condensed
matter physics, because of both the interesting basic physics involved \cite{sato}, and also the possible applications in decoherence-free quantum computing based on the Majorana zero
modes (MZMs) at their ends.\cite{kitaev,nayak,alicea,aasen,lobos,marra}

The simplest model that presents MZMs at the ends is the Kitaev chain for $p$-wave superconductors \cite{kitaev2}.
Lutchyn \textit{et al.} \cite{lutchyn2010} and Oreg \textit{et al.} 
\cite{oreg2010} proposed a model for topological superconducting wires that includes spin-orbit coupling (SOC), proximity-induced $s$-wave superconductivity, and an applied
magnetic field perpendicular to the direction of the SOC. 
The phase diagram of the lattice version of this model has been
calculated recently \cite{diag}.
For reasonable parameters, the model has a topological phase
with MZMs localized at its ends as the Kitaev chain. MZMs of similar wires were found experimentally \cite{wires-exp1,wires-exp2,wires-exp3,wires-exp4}.

A difficulty of these experiments is to identify unambiguously 
that the zero modes are of topological origin, which implies that
they remain at zero energy and localized at the end of the nanowire if small perturbations are applied to the system.
Several authors studied the system consisting of 
a topological superconducting wire and a quantum dot (QD)
\cite{wires-exp4,prada,ptok,clarke,deng,ricco,gru,souto}.
Using the model mentioned above for $s$-wave topological
superconducting wires, Prada \textit{et al.} proposed that 
a QD at the end of the nanowire may be used as a powerful spectroscopic tool to quantify the degree of Majorana nonlocality through a local transport measurement \cite{prada}.
This proposal has been confirmed experimentally \cite{deng}. 
A similar procedure has been also proposed for the Kitaev spinless model \cite{clarke}, and further theoretical studies
have been made recently for the spinfull model \cite{gru}
and a minimal Kitaev chain \cite{souto}.

In general, the energy of the dot level is varied changing the gate potential, and the low-energy levels detected 
by the conductance show either a crossing 
(``bowtie'' shape like in Fig. \ref{compar})
or a “diamond” pattern (like in Fig. \ref{diamond} top)
\cite{prada,souto,ricco}.

Compared to the large amount of theoretical works studying
non-interacting superconducting wires, studies of the 
effects of interactions are rare \cite{ruiz,camja,pandey}.
The effect of nearest-neighbor
repulsion on the Kitaev model has been studied by 
Density Matrix Renormalization Group \cite{tho13,ger16}. The authors find that the quality of the MZMs is not affected by moderate repulsion.
Instead, using approximate methods, it has been found that long-range repulsion spoils the MZMs \cite{wiec}.

Recently it has been pointed out that Coulomb 
repulsion between the electrons of the dot and the 
nanowire might lead to spoiling of the quality of the 
MZMs due to an effective increase of the coupling between
the MZMs localized at the left and at the right of the 
nanowire \cite{ricco}. In particular, Ricco \textit{et al.} considered a spinless model
consisting of a Kitaev chain with a QD at its left end.
There is hopping between the QD and the chain and also
the authors included an interaction 

\begin{equation}
H_V=V n_d n_w,  \label{hv}
\end{equation}
where $n_d$ is the number of electrons in the dot 
and $n_w$ is the total number of electrons in the superconducting wire. In terms of creation and annihilation operators of spinless fermions at each site, it can be written as
$n_w=\sum c_{j}^{\dagger }c_{j}$.
The authors replaced 
this operator by the parity operator at low energies
\begin{equation}
n_w \sim i \gamma_L \gamma_R +1/2  \label{nw}
\end{equation}
(neglecting the 
excited states), 
where $\gamma_\nu$ is the Majorana at the end $\nu$ (left or right) of the wire \textit{at a given chemical potential}. We discuss this approximation in Section \ref{kitac}.
Neglecting the states at higher energy, the authors
estimate the value of $V$, solve exactly the effective
low-energy model and show that $H_V$ contributes to the 
displacement of the MZMs from zero energy, spoiling the 
Majorana quality and the topological properties.

Typically, the low-energy effective Hamiltonian serves as an excellent approximation to the full Hamiltonian. An illustration of this is the quantitative agreement observed in both descriptions for the Josephson current between two topological superconducting wires \cite{tomo}. However, a simple argument suggests
that this might not be the case for the interaction 
given by Eq. (\ref{hv}). If one replaces $V n_d$ by a real number, the resulting term is equivalent to a shift in the chemical potential.  It is well-established that, as long as the system remains in the topological phase, this term does not compromise the integrity of the Majorana modes. However, the operator in Eq. (\ref{nw}) splits them,
revealing a limitation of the effective low-energy model. In the complete Hamiltonian, the states described by  
$\gamma_\nu$ \textit{change their form} and accommodate to the new chemical potential, as expected from the robust nature of topological end states.
This change is not captured by the low-energy effective Hamiltonian. Therefore a study including also the higher-energy states is desirable.

In addition, Eq. (\ref{hv}) assumes the same repulsion to all sites of the chain. However, since the wire has induced superconductivity, there is no 
electric field inside it, and therefore, one expects that 
the Coulomb repulsion between the QD and the end of the wire opposite to it is negligible. In Ref. \cite{wiec} 
the intermediate scenario of a decaying long-range 
repulsion was taken. We contend that it is more realistic to assume a repulsion only between the QD and the first site of the chain. 

\begin{figure} 
 \includegraphics[width=\columnwidth]{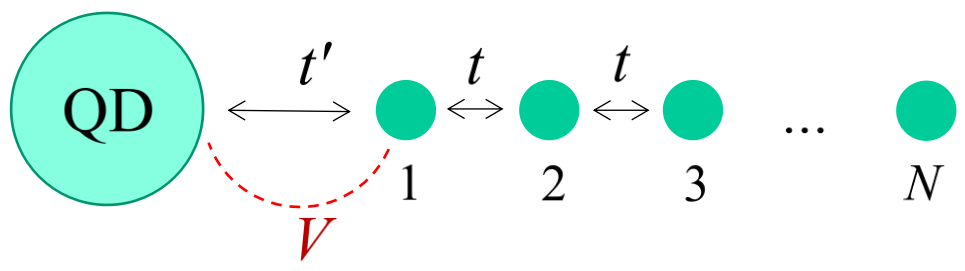}
\caption{(Color online) Scheme of the studied model Eq. (\ref{ham}) representing the interaction $V$ and the hopping terms. There is in addition a $p$-wave superconducting term in the wire [proportional to 
$\Delta$ in Eq. (\ref{ham})].}
\label{scheme}
\end{figure}

In this work we calculate the low-energy spectrum of a Kitaev chain in which the leftmost side has a hopping and also a 
repulsion to a QD state. 
A scheme of the model is presented in Fig. \ref{scheme}.
The repulsion is treated in the unrestricted Hartree-Fock approximation. In Section \ref{model} we describe the model and the approximation.  In Section \ref{res} we show the numerical results.
Section \ref{sum} contains a summary and discussion.

\section{Model and approximation}

\label{model}

The Hamiltonian of the Kitaev chain interacting with a QD is

\begin{eqnarray}
H &=&\sum_{j=1}^{N-1}(-tc_{j+1}^{\dagger }c_{j}+\Delta c_{j+1}^{\dagger
}c_{j}^{\dagger }+\mathrm{H.c.})-\mu \sum_{j=1}^{N}c_{j}^{\dagger }c_{j} 
\notag \\
&&+\epsilon_d d^{\dagger }d-t^\prime (d^{\dagger }c_{1}+\mathrm{H.c.}) \notag \\
&&+V\left( n_d  -\frac{1}{2} \right) 
\left( n_1 -\frac{1}{2} \right),  \label{ham}
\end{eqnarray}
where $n_{d}=d^{\dagger }d$ and $n_{1}=c_{1}^{\dagger }c_{1}$. The first two
terms of Eq. (\ref{ham}) describe the Kitaev chain with hopping $t$, $p$-wave
superconducting order parameter $\Delta $ and chemical potential $\mu $. The
third term describes the QD. The fourth term is the hopping between the QD
and the Kitaev chain and the last term is the Coulomb repulsion between the
electrons in the QD and the ones at the leftmost site of the chain. We treat
this term in the unrestricted Hartree-Fock approximation:

\begin{eqnarray}
n_{d}n_{1} &\simeq &\left\langle n_{d}\right\rangle n_{1}+n_{d}\left\langle
n_{1}\right\rangle -\left\langle n_{d}\right\rangle \left\langle
n_{1}\right\rangle   \notag \\
&&-\left\langle d^{\dagger }c_{1}\right\rangle c_{1}^{\dagger }d-d^{\dagger
}c_{1}\left\langle c_{1}^{\dagger }d\right\rangle +\left\langle d^{\dagger
}c_{1}\right\rangle \left\langle c_{1}^{\dagger }d\right\rangle   \notag \\
&&+\left\langle d^{\dagger }c_{1}^{\dagger }\right\rangle c_{1}d+d^{\dagger
}c_{1}^{\dagger }\left\langle c_{1}d\right\rangle -\left\langle d^{\dagger
}c_{1}^{\dagger }\right\rangle \left\langle c_{1}d\right\rangle .  \label{hf}
\end{eqnarray}

We note that our model is different from that of Ricco \textit{et al.}, who considered repulsion with all the sites of the wire with the 
same intensity \cite{ricco}.
Another difference is the treatment of repulsion. 
Ricco \textit{et al.} treated the repulsion
exactly in an effective model within a low-energy 
subspace \cite{ricco}.
We include all states but treat the repulsion using the 
approximation of Eq. (\ref{hf}).

\section{Results}

\label{res}

We take $t=1$ as the unit of energy and choose 
$\Delta=0.2$. 

\subsection{Isolated Kitaev chain}
\label{kitac}

\begin{figure}[th]
\begin{center}
\includegraphics*[width=0.9\columnwidth]{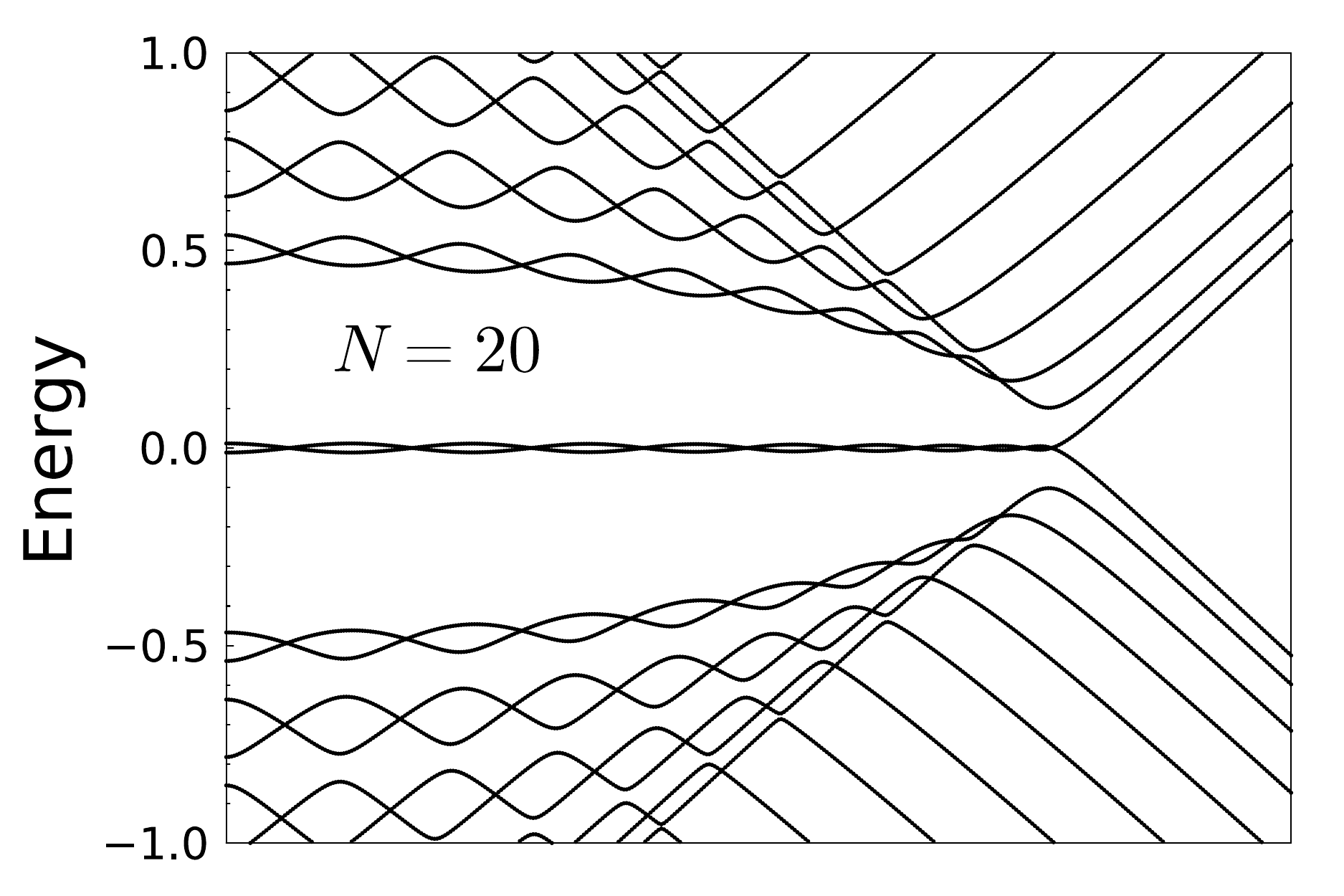}\\
\vspace{-0.2cm}
\includegraphics*[width=0.9\columnwidth]{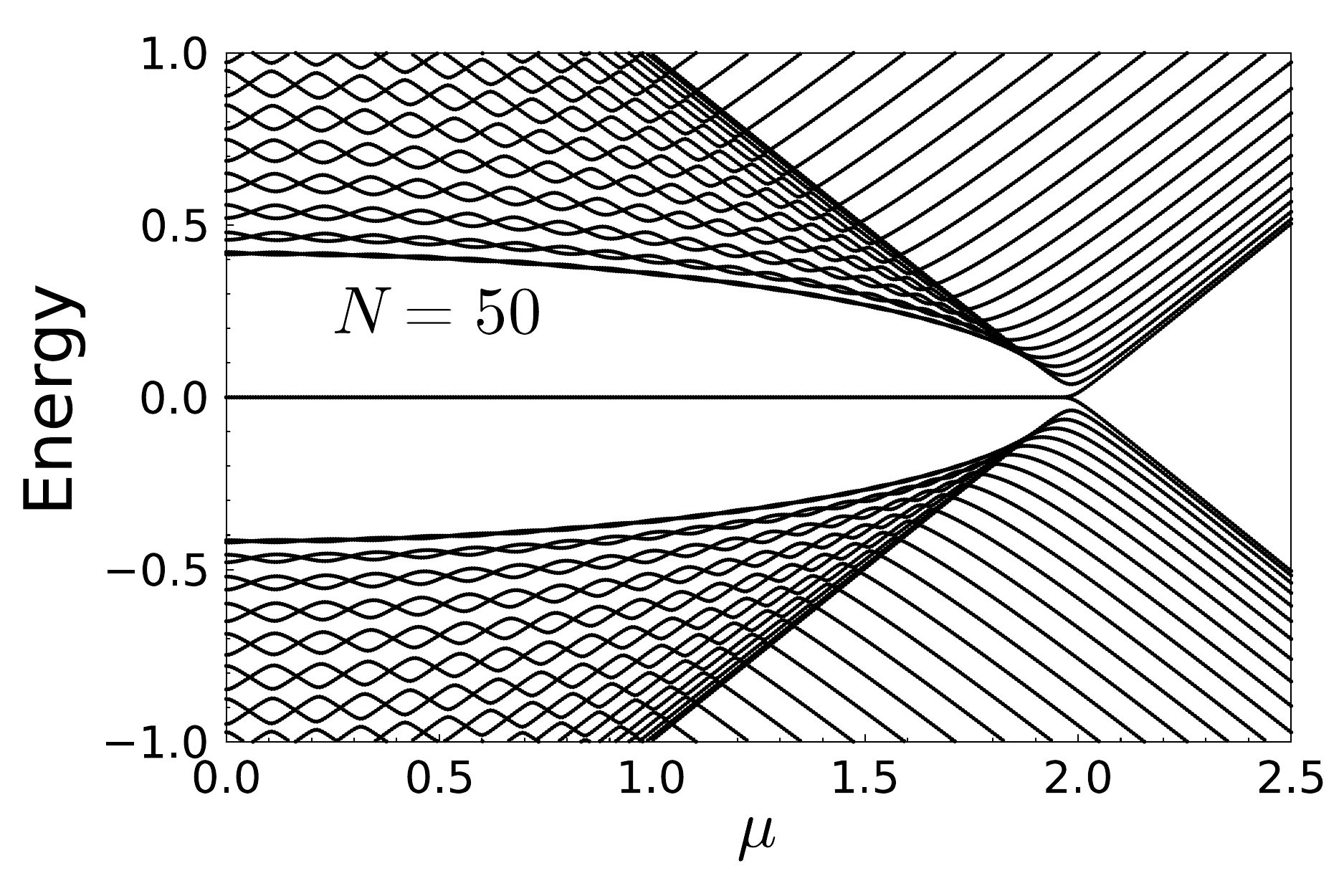}\\
\includegraphics*[width=0.8\columnwidth]{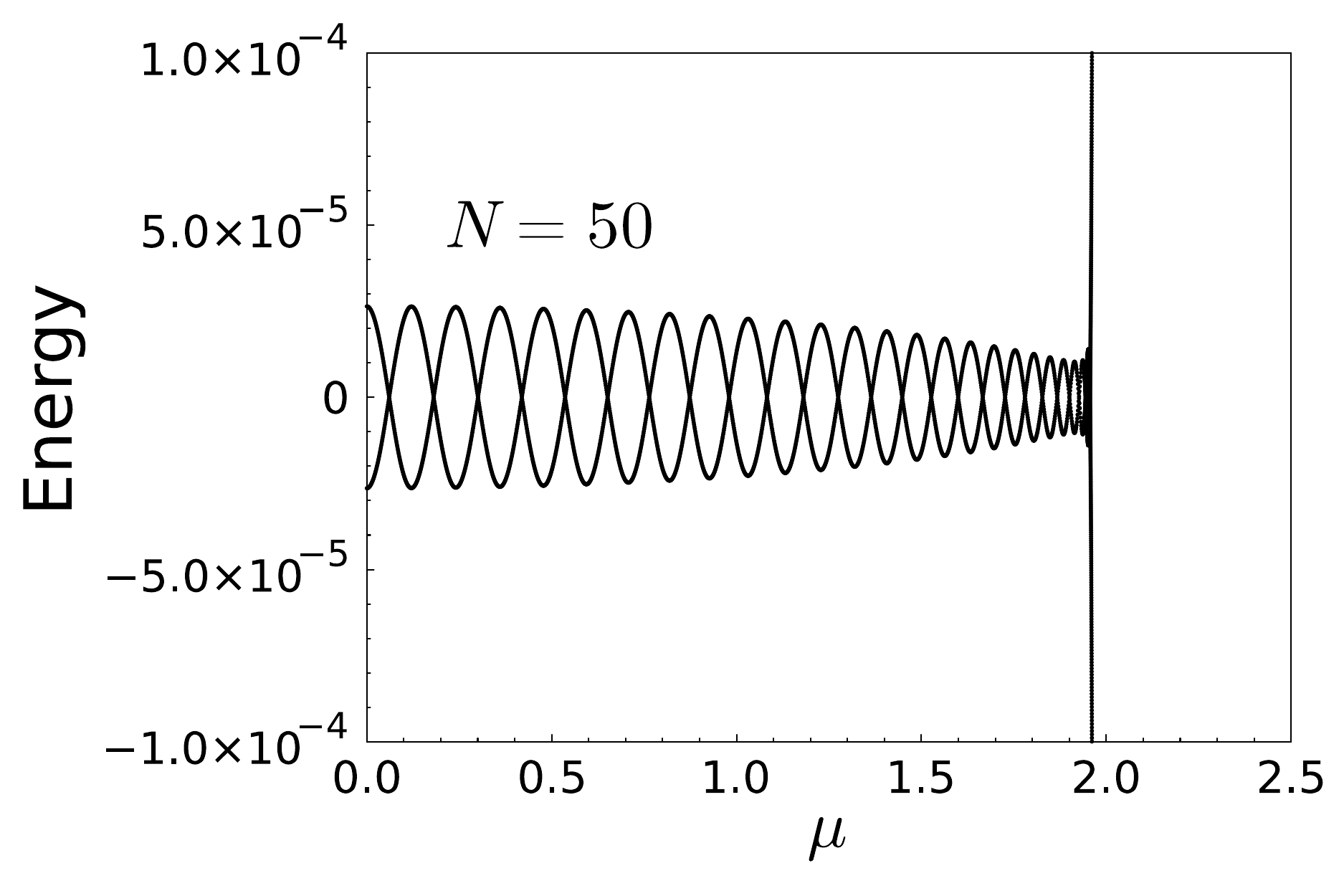}
\end{center}
\caption{Eigenvalues of the Kitaev chain 
for 20 sites (top) and 50 sites (middle and bottom) 
as a function of the chemical potential.
Other parameters are $t=1$, $\Delta=0.2$. }
\label{kita}
\end{figure}

For later comparison, we discuss first the isolated Kitaev chain (without the quantum dot) 
for two different lengths of the 
chain. 
The resulting energies are shown in Fig. \ref{kita} as a function of the chemical potential $\mu$. The curve is symmetric under the change of sign of $\mu$, and 
therefore only positive values of $\mu$ are displayed
in the figure.
As it is known, the system is topological for 
$|\mu|< 2 t$. In this region, there are two low-energy
states at energies near zero. For the infinite chain, these states 
correspond to the left and right MZMs $\gamma_L$ and 
$\gamma_R$ localized at the ends of the chain. 
In a finite chain, these modes are mixed with an 
effective term $\lambda i \gamma_L \gamma_R$ and the energies are split in $\pm \lambda$. As expected, $\lambda$ decays exponentially with increasing system size. From Fig. \ref{kita}, one can see that $\lambda$ decreases almost 
four orders of magnitude when the length of the chain is increased from 20 to 50 sites.

One can also see from the figures that $\lambda$ oscillates as the chemical potential is varied. The period 
of oscillation is more that two times smaller for 50 sites 
in comparison with 20 sites, and it is also smaller for larger $|\mu|$, 
near the topological transition to the trivial phase.

\subsubsection{Limitations of simple low-energy effective models}
\label{limi}

We have also investigated the low-energy part of 
$n_w=\sum c_{j}^{\dagger }c_{j}$ 
and $n_1=c_{1}^{\dagger }c_{1}$ in the topological phase. Clearly after changing
the basis to the eigenstates of the chain, and taking the lowest-energy part of the operator, one has
$n_w \sim Af^{\dagger }f+Bff^{\dagger }$, where $A$ and $B$ are positive real numbers and
$f^{\dagger }$ is the eigenstate of lowest positive 
energy ($[H,f^{\dagger }]=|\lambda|f^{\dagger }$).
Except for an irrelevant phase, 
it corresponds to 
$f^{\dagger }=[\gamma_L -\text{sign}(\lambda) i \gamma_R]/2$. 

The electron-hole symmetry of the isolated Kitaev chain $c_{j}^{\dagger } \rightarrow (-1)^jc_{j}$, transforms $H(\mu)$ into $H(-\mu)$ and 
$f^{\dagger }f$ into $1-f^{\dagger }f$. Therefore using
the form $n_w \sim Cf^{\dagger }f+B$, with $C=A-B$, 
$C$ should be an odd function of $\mu$ and for 
$\mu=0$, $A=B=1/2$ and $C=0$,
as we confirm numerically.
This result can be analytically derived in a straightforward manner within the exactly solvable limit 
($|\Delta|-|t|=\mu=0$) of the Kitaev chain.
For other values of $\mu$ 
inside the topological phase, we find that $B \sim 1/2$ 
and $C$ is of the order of 0.01 (0.001 or less) for 
$N=20$ ($N=50$). This factor $C$ is lacking in the approximation of Eq. (\ref{nw}) taken by Ricco \textit{et al.} \cite{ricco}
and therefore the effect of Coulomb repulsion on
splitting the MZMs is exaggerated in their work.

Similarly, the low-energy part of $n_1$ containing 
two $f$ operators has the form 
$n_1 \sim C_1 f^{\dagger }f+B_1$, with $C_1=0$ for $\mu=0$.
For other values of $\mu$, the order of magnitude of 
$C=1$ is 0.01 ($10^{-5}$) for $N=20$ ($N=50$).

The findings presented in the subsequent section, especially for short chains and when $\mu \sim 0$
showing a significant splitting of the low-energy modes (see Fig. \ref{diamond}), indicate that the interaction of the mode $f$ with higher energy modes is important. 
As above, the coefficient of $f^{\dagger }f$ should vanish by symmetry for $\epsilon_d=V=\mu=0$ but a splitting 
is present in Fig. \ref{diamond}.
In such cases, particularly when the low-energy modes are not well confined to the ends, the adequacy of the simplest effective model becomes questionable in accurately characterizing the quality of the Majorana zero modes (MZMs).
The inclusion of higher energy states has been addressed perturbatively in the examination of boundary states in two-dimensional topological superconductors \cite{rodri}.

In addition, the adjustment of the MZMs to a new configuration, remaining near zero energy, particularly when $\mu$ is varied within the topological phase, implies a necessary mixing of low-energy modes (described by $f$) with higher energy counterparts. Otherwise, the splitting would increase linearly with the change in $\mu$, contrary to what is shown in Fig. \ref{kita}.

\subsection{Effect of the quantum dot}
\label{kitadot}

\begin{figure}[th]
\begin{center}
\includegraphics*[width=0.9\columnwidth]{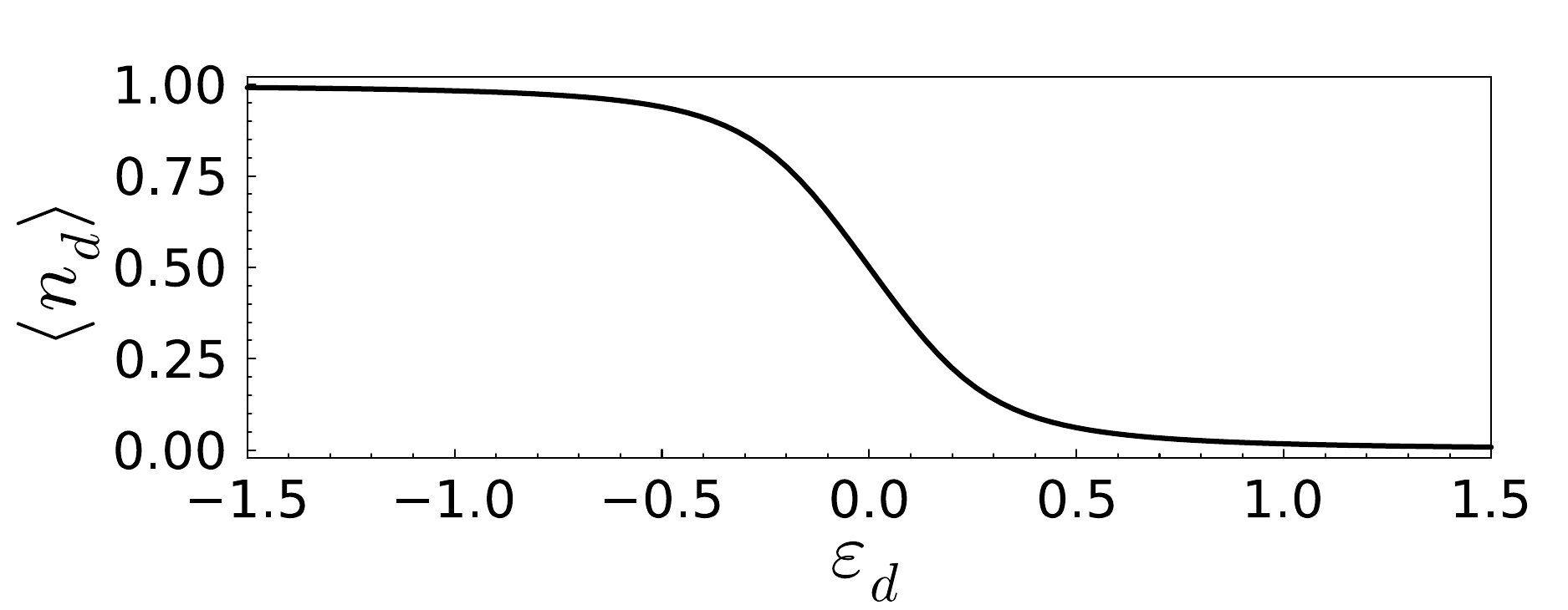}\\
\vspace{-0.45cm}
\includegraphics*[width=0.9\columnwidth]{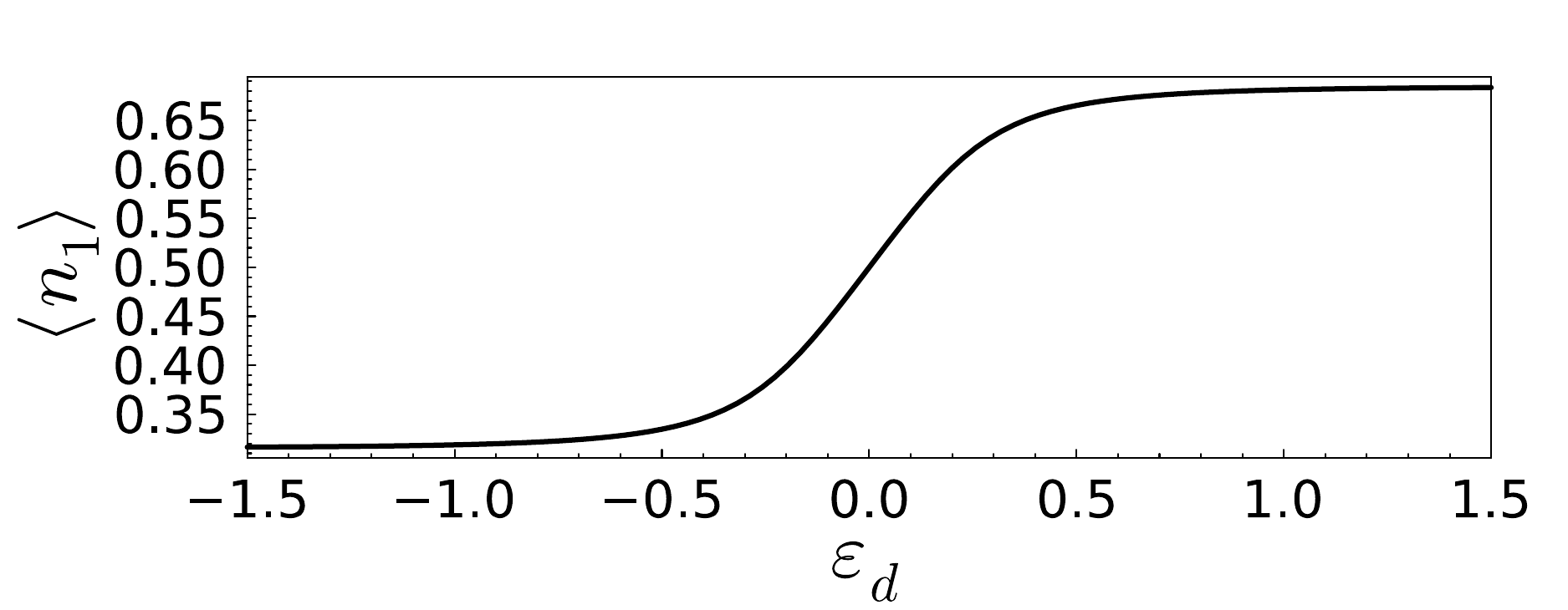}\\
\vspace{-0.45cm}
\includegraphics*[width=0.9\columnwidth]{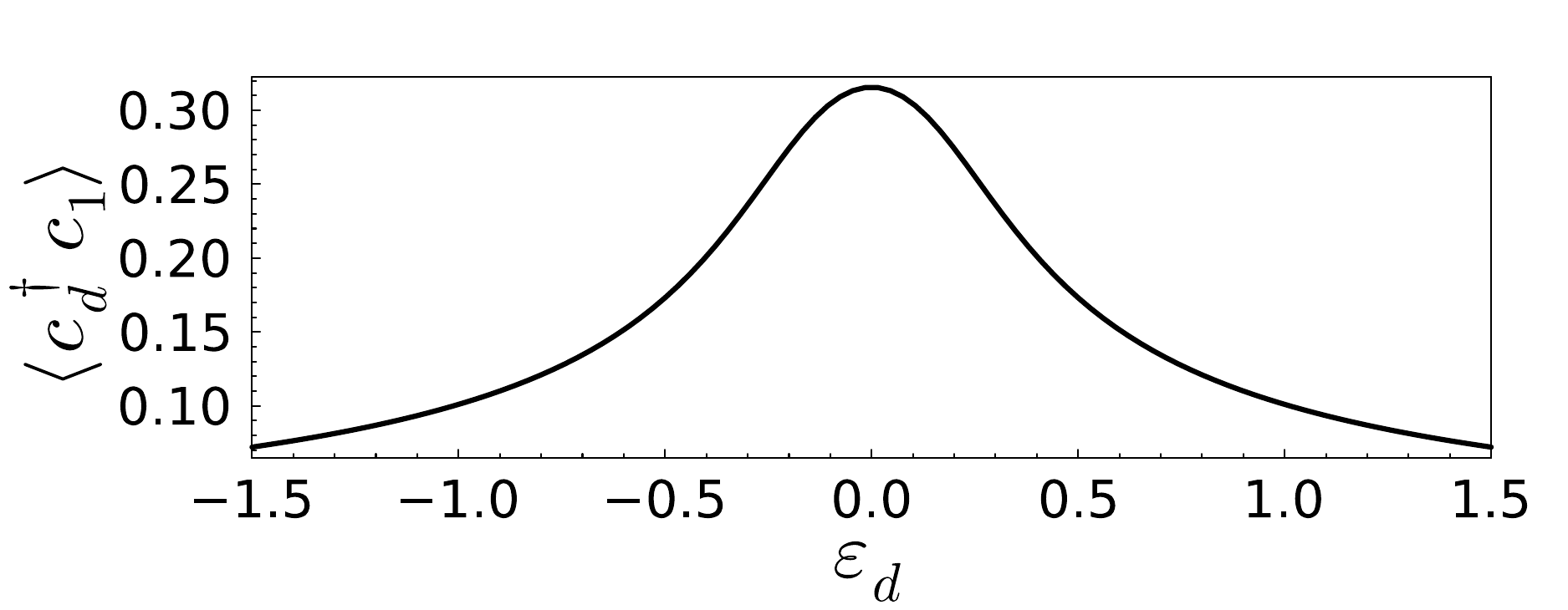}\\
\vspace{-0.45cm}
\includegraphics*[width=0.9\columnwidth]{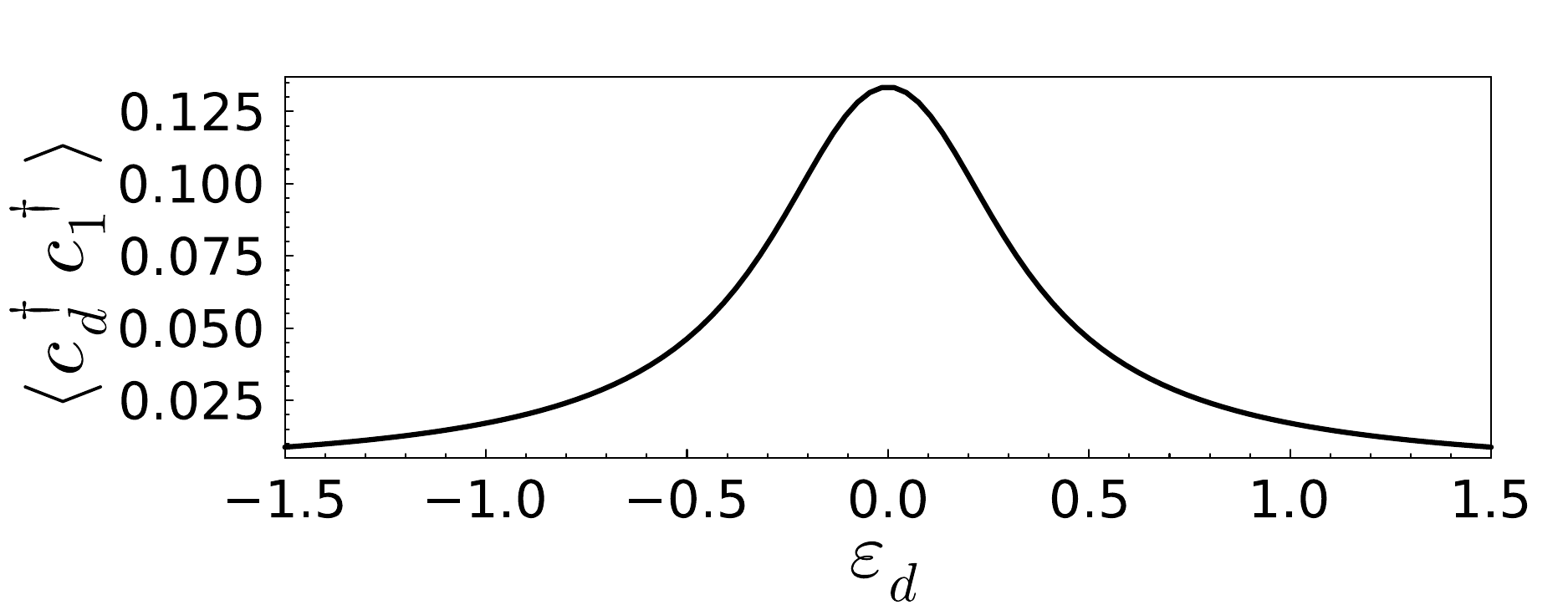}
\end{center}
\caption{Expectation values entering Eq. (\ref{hf}) as a function of dot energy. 
Other parameters are $N=50$, $t=V=1$, 
$\Delta=t^\prime=0.2$, and $\mu=0$.}
\label{self}
\end{figure}

Including the QD and to facilitate the discussion of the effects of the nearest-neighbor repulsion $V$, we represent in Fig. \ref{self}, the expectation values that enter in the unrestricted Hartree-Fock approximation, Eq. (\ref{hf}), determined selfconsistently. 
We have chosen $t^\prime=0.2$,
$V=1$, $\mu=0$ and a chain of 50 sites excluding the quantum dot.
The results are rather insensitive to system size.

As expected, the occupancy of the dot is near 1 when its energy is negative and large in magnitude compared to
$t^\prime$ ($-\epsilon_d \gg t^\prime$), it is equal to
1/2 for $\epsilon_d=0$ and it is near 0 for $\epsilon_d \gg t^\prime$. 

In contrast, the occupancy of the first site of the dot 
$\langle n_{1} \rangle$ follows qualitatively the opposite behavior: 
when $\langle n_d \rangle > 1/2$, the first site feels the repulsion with the electrons in the dot and its occupancy decreases, but its hopping with the rest of the chain
moderates this effect and the occupancy deviates from 0.5 
in less than 0.2.

The expectation value of the hopping 
$\langle d^{\dagger } c_{1} \rangle$
follows qualitatively the behavior expected for a 
diatomic heteronuclear molecule, with a single orbital per atom, in which the two atomic states are hybridized. 
The expectation value is maximum when both atomic levels coincide ($\epsilon_d=0$) and decreases symmetrically
with the difference between atomic levels. 
The half width of
the curve is expected to be of the order of the effective hopping, which in this case is 
$t^\prime_\text{eff}= t^\prime + V \langle d^{\dagger } c_{1} \rangle$. For
$\epsilon_d=0$, this value is near 0.5, consistent with the resulting half width and more that two times larger 
than the bare value $t^\prime=0.2$.

The pairing expectation value 
$\langle d^{\dagger } c_{1}^{\dagger } \rangle$
follows qualitatively a similar dependence with
the dot energy as the hopping contribution discussed above, 
but with smaller values. Its dependence with
$\epsilon_d$ is also narrower. Its physical origin is
a proximity induced $p$-wave superconductivity,
which is larger when the energy of the dot is nearer to the chemical potential of the wire.

\begin{figure}[th]
\begin{center}
\includegraphics*[width=0.9\columnwidth]{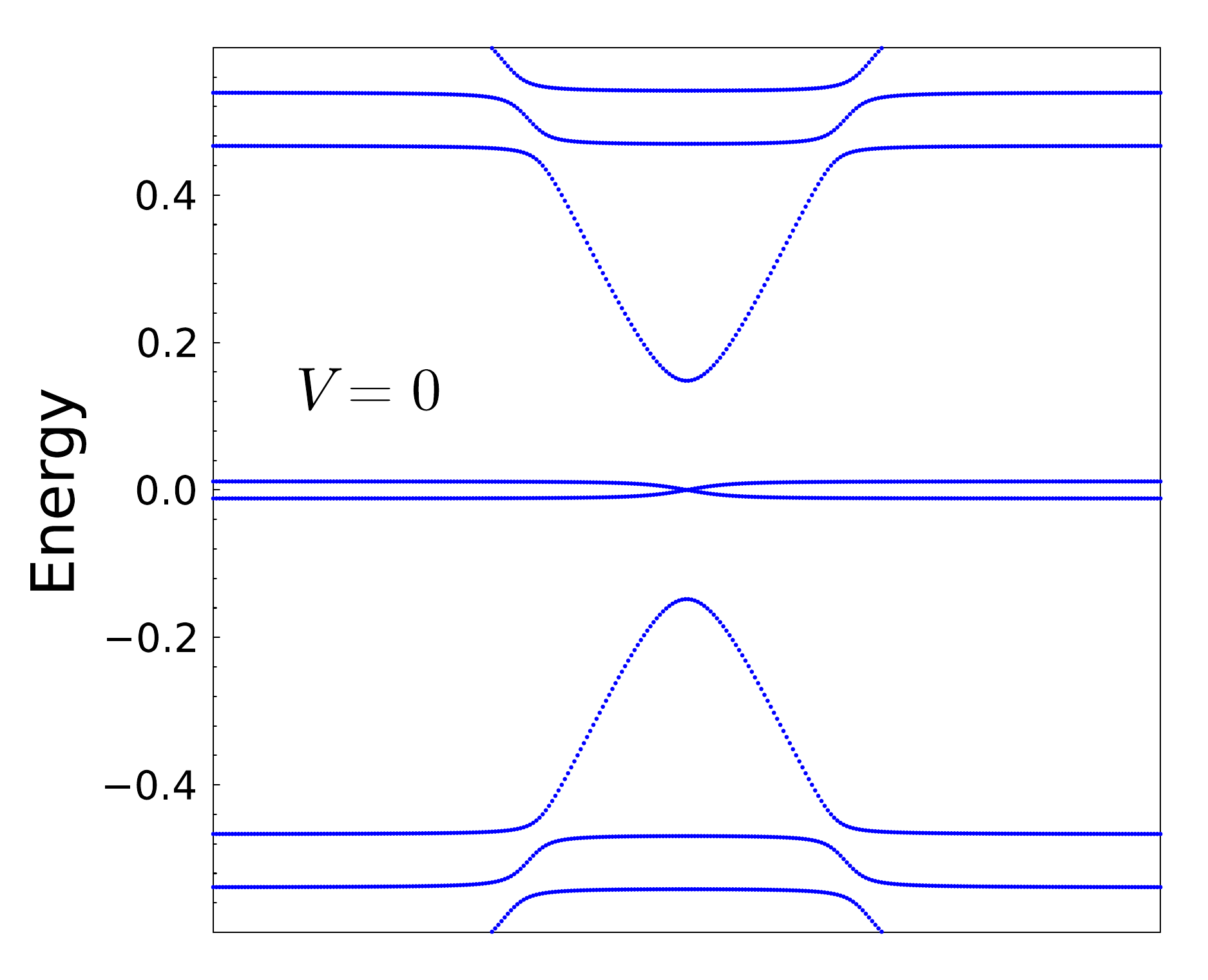}\\
\vspace{-0.3cm}
\includegraphics*[width=0.9\columnwidth]{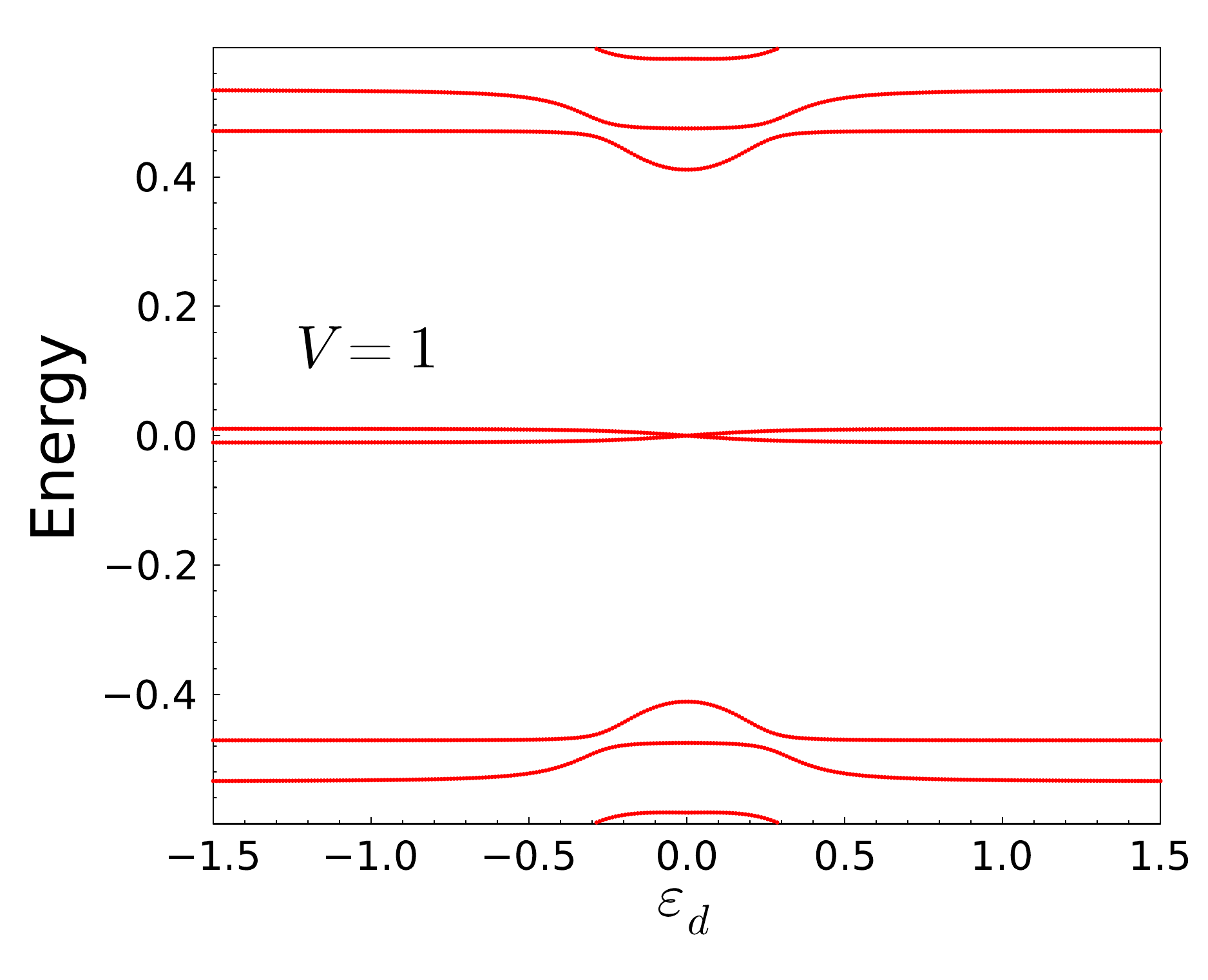}
\vspace{-1.0cm}
\includegraphics*[width=0.8\columnwidth]{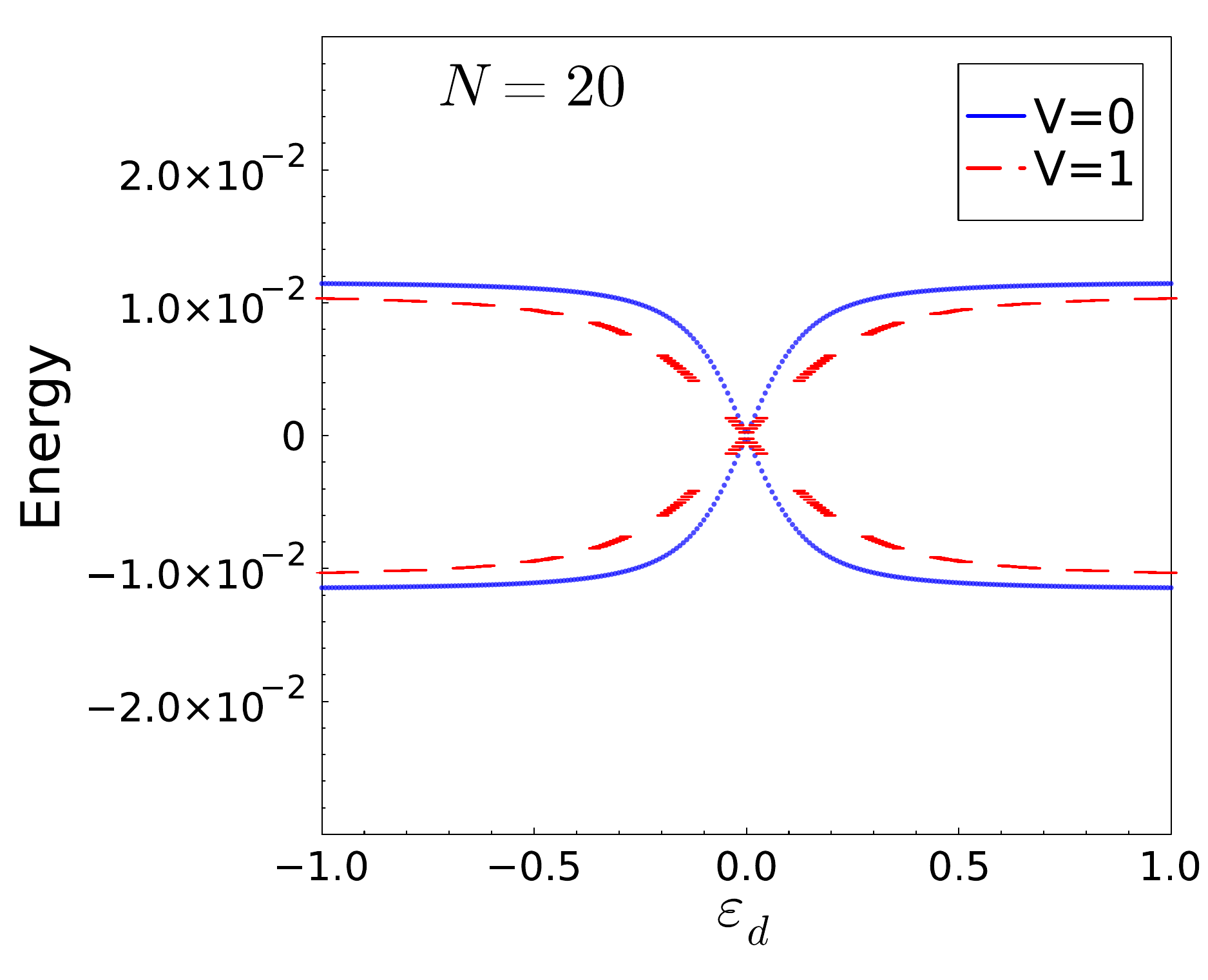}
\end{center}
\caption{(Color online) Energies of the system
as a function of the energy of the dot for two values of $V$.
Other parameters are $N=20$, $t=1$, 
$\Delta=t^\prime=0.2$, and $\mu=0$.}
\label{ene20}
\end{figure}

\begin{figure}[th]
\begin{center}
\includegraphics*[width=0.9\columnwidth]{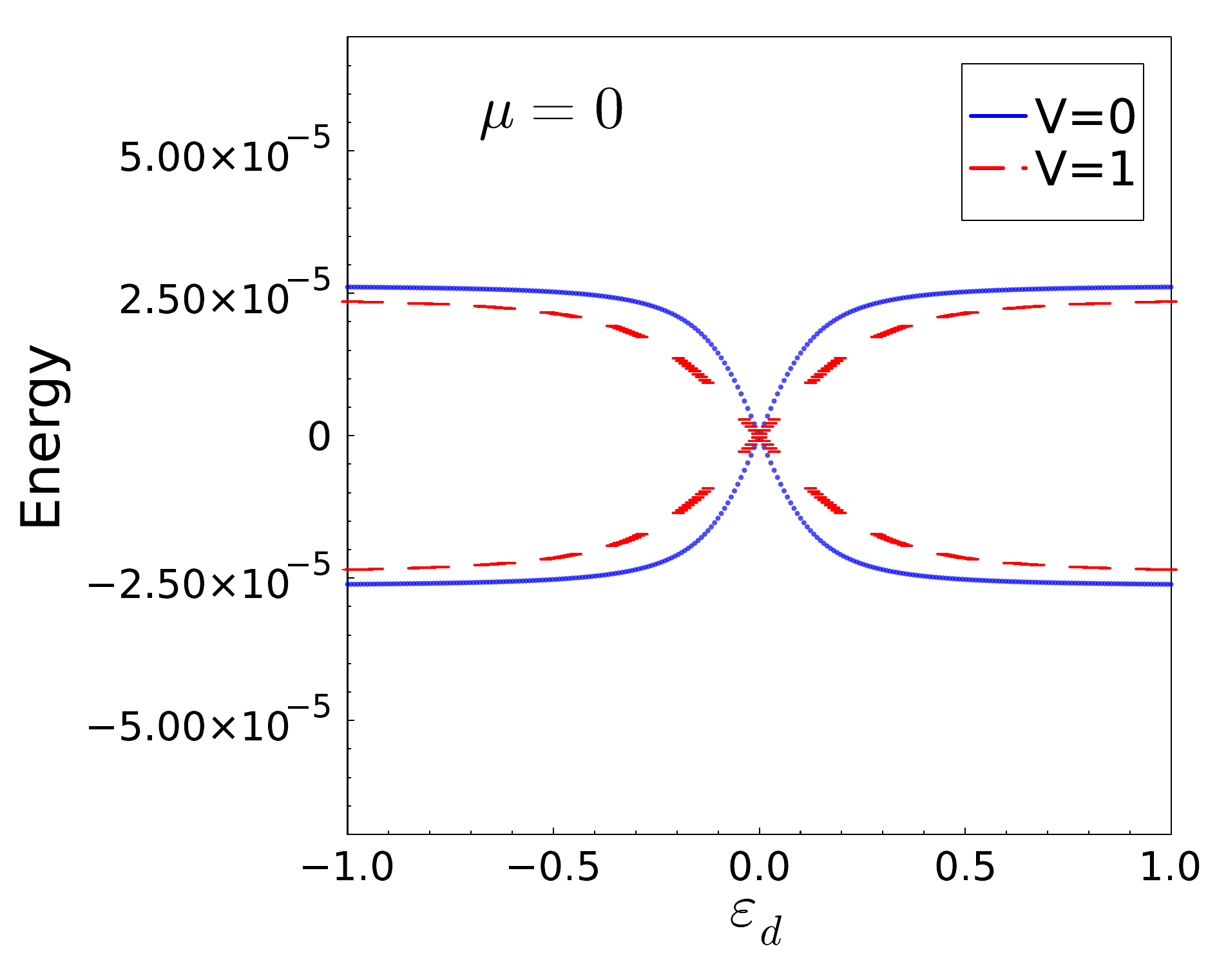}\\
\vspace{-0.9cm}
\includegraphics*[width=0.9\columnwidth]{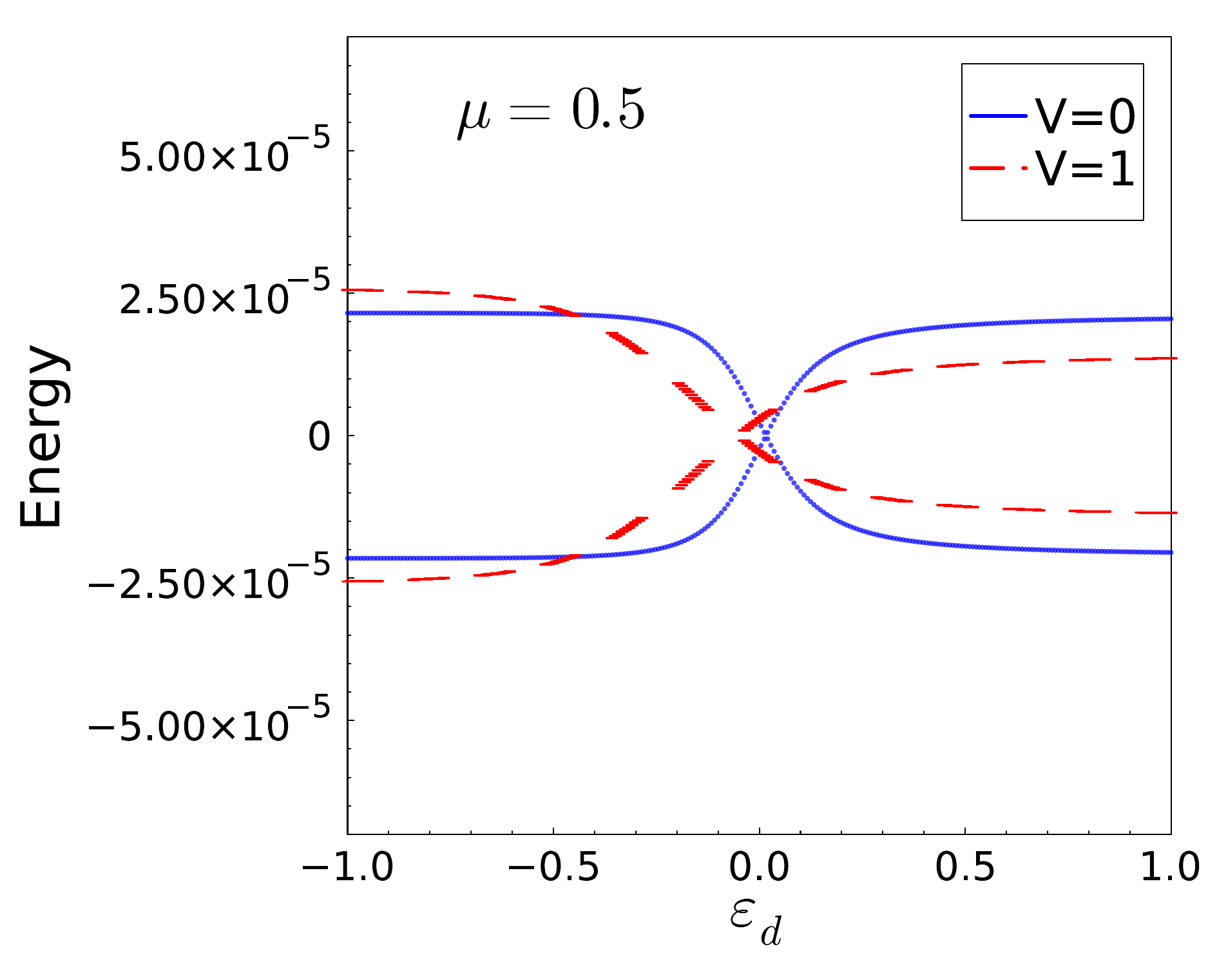}
\end{center}
\caption{(Color online) Comparison of the 
two energies nearer to zero in the system
as a function of the energy of the dot between $V=0$ and $V=1$
for $\mu=0$ (top) and $\mu=0.5$ (bottom).
Other parameters are $N=50$, $t=1$, and 
$\Delta=t^\prime=0.2$.}
\label{compar}
\end{figure}

The resulting eigenvalues of the system 
as a function of the dot energy for $V=0$ and $V=1$ 
are compared in Fig. \ref{ene20} for a chain of 20 sites.
For 50 sites the discussion below is practically the same,
but the results are displayed more clearly in the smaller system. For the sake of brevity we omit displaying the results for 50 sites except near zero energy (Fig. \ref{compar}).
We discuss first the case $V=0$. 
For large $|\epsilon_d|$, the eigenvalues at small energies (of absolute value less than 1) are 
practically the same as those of the isolated Kitaev chain
shown in Fig. \ref{kita}. For $\mu=0$, the results are symmetric under interchange of the sign of 
$\epsilon_d$.
In addition to the states of the isolated chain, 
there are roughly speaking two other symmetric states at energies $\pm E_m$,
which at a first approximation corresponds to that of higher
absolute value 
of an heteronuclear molecule (as that mentioned above)
that mixes two states with energies $\epsilon_d$ and zero.
For large $|\epsilon_d|$, $E_m \sim \epsilon_d$ and for 
$\epsilon_d=0$, $E_m \sim t^\prime$. These states actually
hybridize with the states of the isolated Kitaev chain 
showing several anticrossings that are evident in Fig. \ref{ene20}.

When $V$ is included, the eigenergies (except the two nearer to zero) 
are modified, particularly those related with the mixing of the dot state near $\epsilon_d=0$. Since as explained above, the effective hopping between the dot and the first state of the chain increases from $t^\prime=0.2$ to $t^\prime \sim 0.5$,
when $V$ is increased from 0 to 1, a similar change takes place for the energies that are near $\pm t^\prime$ in Fig.
\ref{ene20}. However, the two energies with lowest absolute value, related with the splitting of the MZMs, are very little modified by $V$.  

In Fig. \ref{compar} we display the energies related to the 
MZMs for two values of $\mu$ and a chain of 50 sites. 
One can see that for $\mu=0$, the inclusion of nearest-neighbor repulsion $V$, at least within our unrestricted Hartree-Fock approximation slightly \textit{decreases} the splitting of the two low-energy states, indicating that the quality of the MZMs actually is \textit{improved} when the repulsion is added. 
For $\mu \neq 0$, the symmetry under change of sign of 
the dot energy $\epsilon_d$ is lost, and the asymmetry is increased with $V$. In any case, the effect of $V$ on the quality of the MZMs remains very small. The shape of the 
curve is similar to that found in previous experiments \cite{deng} and theory \cite{prada,ricco}.

\begin{figure}[hb]
\begin{center}
\includegraphics*[width=0.90\columnwidth]{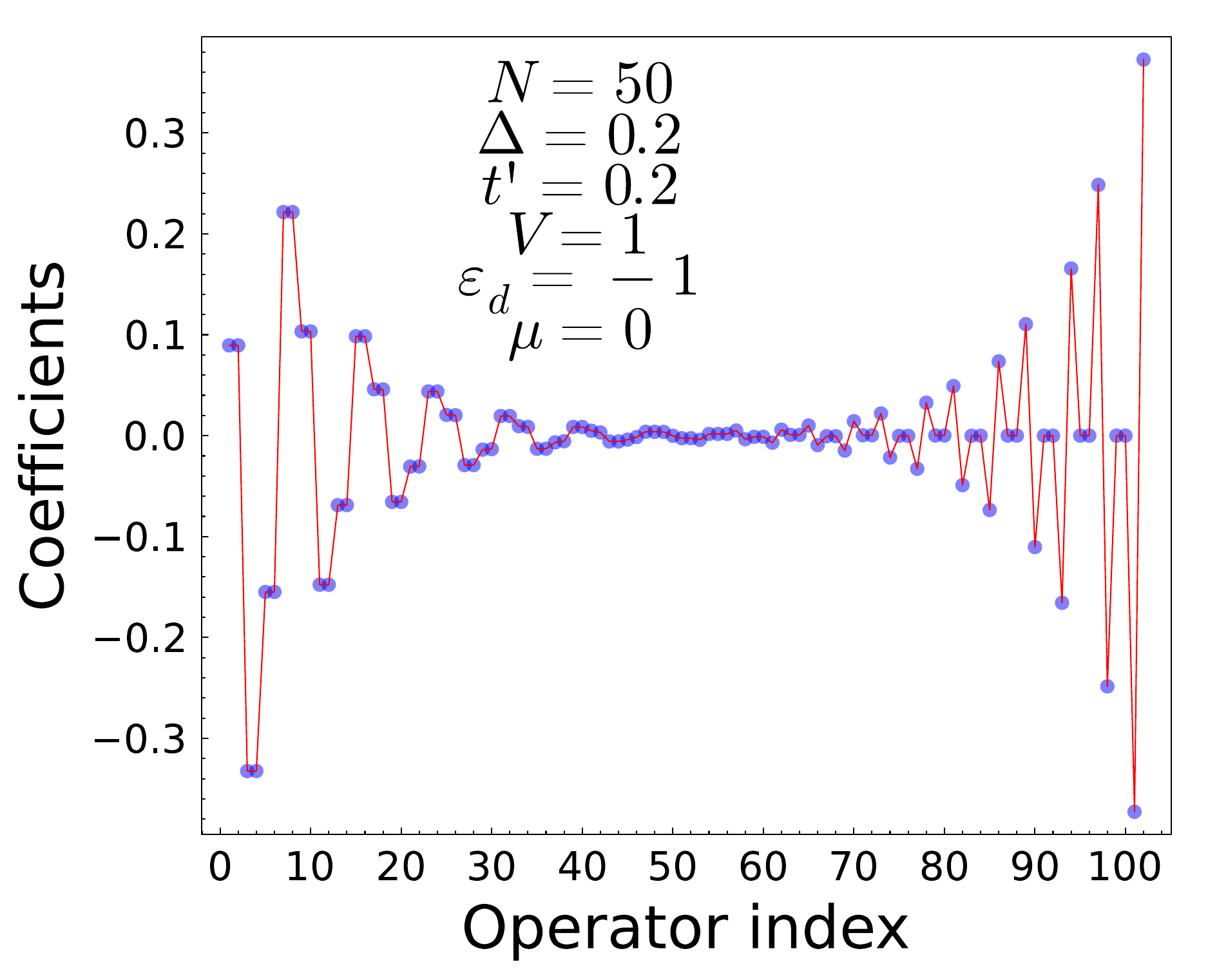}
\end{center}
\caption{(Color online) Coefficients of the lowest eigenstate of the system.
Parameters are $N=50$, $t=V=1$, 
$\Delta=t^\prime=0.2$, $\epsilon_d=-1$ and $\mu=0$.}
\label{wei}
\end{figure}

In Fig. \ref{wei}, we show the coefficients of the lowest eigenstate with positive energy for the parameters indicated inside the figure. The fermion is written as 
$\sum_i \alpha_i f_i$, where $\alpha_i$ are the 102 coefficients and the order of the corresponding fermions $f_i$ 
is $f_1=d^\dagger$, $f_2=d$, $f_3=c_1^\dagger$, $f_4=c_1$, ...
$f_{102}=c_{50}$. As expected, the state is a mixture of the MZMs at the ends with negligible weight in the middle of the chain. However, in contrast to the isolated Kitaev chain, 
there is a significant weight of the state also at the dot, with a probability which is about $1/10$ compared to that of the first site in the chain. This probability increases with decreasing $|\epsilon_d|$.

\begin{figure}[th]
\begin{center}
\includegraphics*[width=0.8\columnwidth]{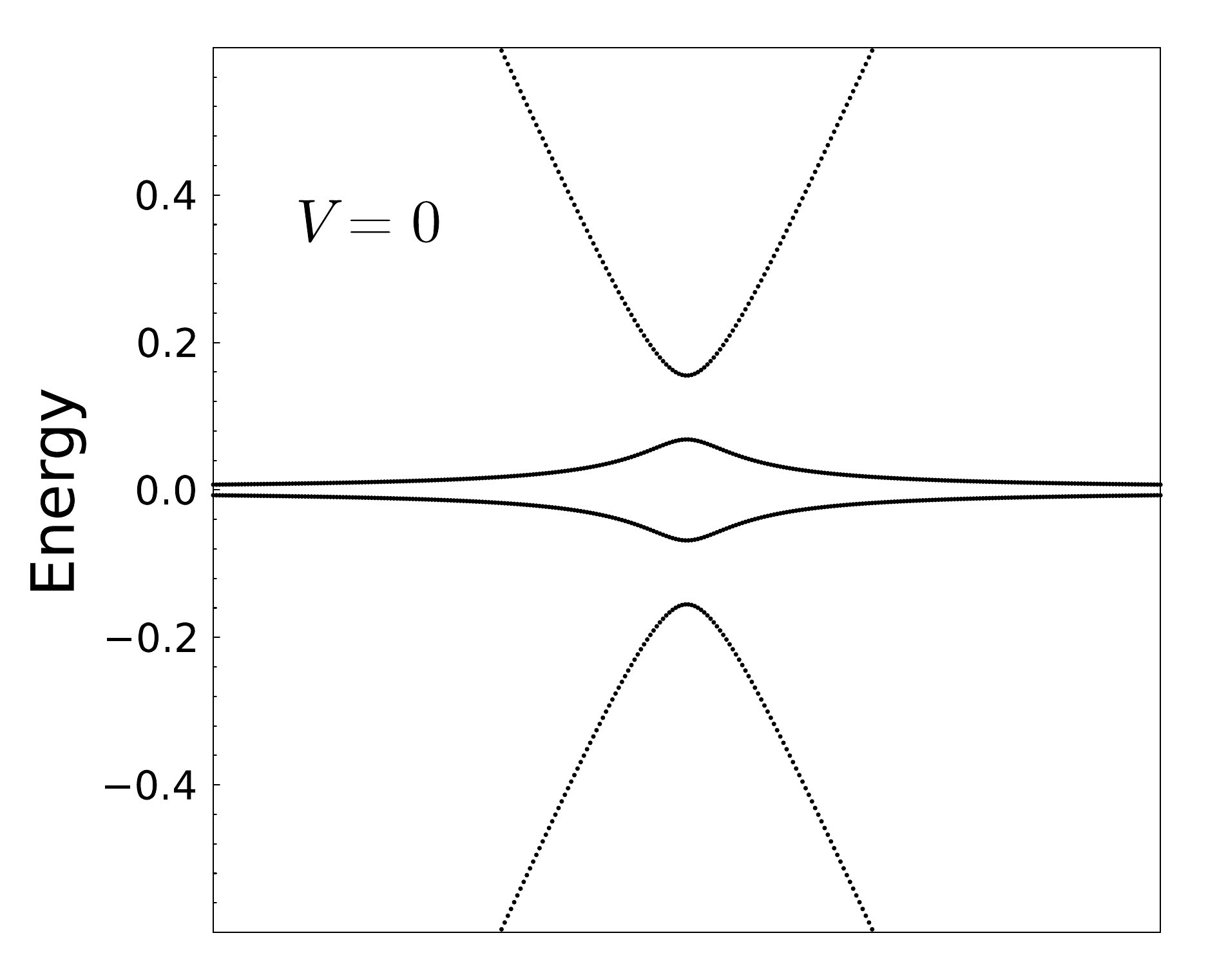}\\
\vspace{-0.3cm}
\includegraphics*[width=0.8\columnwidth]{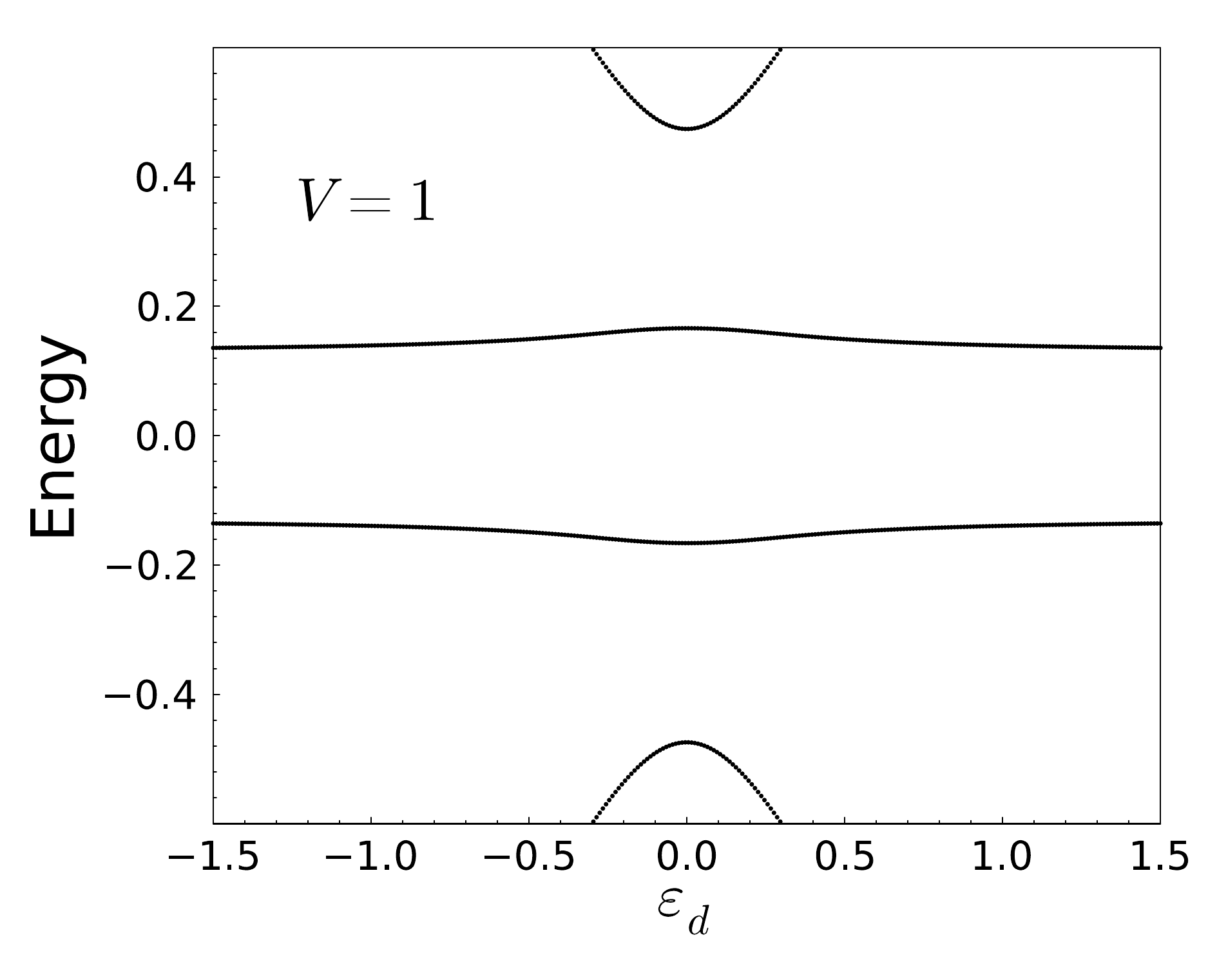}
\end{center}
\caption{Energies of the system
as a function of the energy of the dot for $V=0$ (top)
and $V=1$ (bottom).
Other parameters are $N=5$, $t=1$, 
$\Delta=t^\prime=0.2$, and $\mu=0$.}
\label{diamond}
\end{figure}

Finally in Fig. \ref{diamond} we display the energies 
for a short chain of 5 sites, 
with a significant mixing of both
MZMs at the ends of the chain. In this case, the weight
of the MZM at the right end is significant at the left end,
and therefore it also feels the repulsion with the quantum dot. For $V=0$ the shape is characteristic of the ``diamond''
one observed in experiment \cite{deng} and in calculations
\cite{prada,ricco} when the hopping between the quantum dot and the MZM at the right end $\gamma_R$ is important \cite{prada,ricco}.

In contrast to the previous cases, now the effect of adding the Coulomb repulsion is significant, leading 
to a strong further splitting of the MZMs, of the order of a fraction of $t^\prime$.

\section{Summary and discussion}

\label{sum}

We have solved a model for a Kitaev chain involving a Kitaev chain on a lattice, linked to a quantum dot at one of its ends through a hopping term and subject to Coulomb repulsion between the relevant state of the quantum dot and the terminal site of the chain. In realistic semiconductor-superconductor heterostructures, crucial for creating Majorana zero modes (MZMs), short-range repulsions are anticipated to play a pivotal role. However, theoretical investigations into the effects of these interactions are scarce.

As the energy of the state of the quantum dot is varied, the energies of the two eigenstates of the system closest to zero exhibit one of the characteristic shapes observed in experiments and prior theories, signaling the presence of Majorana zero modes (MZMs) at the ends of the wire, coupled between them. In one of them (``bowtie'' shape), the energies of the two states cross when the energy of the quantum dot is near to the Fermi energy. 
In this case, the coupling between MZMs is weak and analyzing the wave function of these eigenstates, one sees that a MZM has a substantial weight at the quantum dot.
Employing the unrestricted Hartree-Fock approximation to account for interatomic Coulomb repulsion, we determine that this repulsion does not significantly impact the quality of the MZMs.

In contrast in the alternative scenario, in which the energies of 
the low-lying states as a function of the dot level have a 
``diamond'' shape, signaling a stronger coupling between the MZMs (shorter chains), the effect of the interatomic
Coulomb repulsion is significant splitting further the MZMs.
In this case, the wave function of the MZM at the opposite end of the dot extends to the site adjacent to the dot and therefore it fills a significant repulsion with the dot. 
Consequently, this outcome aligns with the observations in Ref. \cite{wiec}, highlighting that interactions between particles situated at more distant sites exert a more destructive influence than interactions between nearest neighbors. In addition, the latter have only
a minor influence on the MZMs of the isolated Kitaev chain \cite{tho13,ger16}.

We believe that our conclusions are also valid 
for more realistic models for the topological
superconductors, such as that of Refs. \cite{lutchyn2010,oreg2010}, even
including additional sub-bands and the
orbital effect of the magnetic field \cite{morteza}.

If the experimental observation of a "diamond" shape in the low-energy spectrum as a function of dot energy, as documented in Ref. \cite{deng}, occurs, it implies that MZMs have a substantial extension across the entire wire. In such cases, Coulomb repulsion at one end exacerbates the degradation of Majorana quality. Conversely, Coulomb repulsion at one end has no discernible impact on well-localized MZMs characteristic of the 
"bowtie" shape.

To summarize the aforementioned results, in situations where the mixing between MZMs in an isolated wire is weak, as expected in systems with well-defined MZMs, the repulsion between the quantum dot and the end of the wire does not compromise the quality of the MZMs. This is a noteworthy and positive outcome for the utilization of quantum dots as indicators of Majorana quality.

\section*{Acknowledgments}

R. K. T. P. has a scholarship of Instituto Balseiro.
A. A. A. acknowledges financial support provided by PICT 2018-01546 and PICT 2020A-03661 of the Agencia I+D+i, Argentina.

\end{document}